\documentclass[reprint,superscriptaddress,amsmath,amssymb,aps,prl,floatfix]{revtex4-1}
\usepackage{graphicx}
\usepackage{bm}
\usepackage{hyperref}
\usepackage{todonotes}
\usepackage{nicefrac}
\usepackage{bbm}
\usepackage[normalem]{ulem}
\usepackage{booktabs}
\usepackage{subfigure}
\usepackage{rotating}
\usepackage[switch,columnwise]{lineno}
\definecolor{fj_color}{cmyk}{1, 0.3, 0, 0}
\definecolor{cwh_color}{cmyk}{0, 0.8, 0.8, 0}
\definecolor{hr_color}{rgb}{0.0, 0.53, 0.74}

\begin{document}

\title{Visualizing the Localized Electrons of a Kagome Flat Band}

\author{Caiyun Chen}
\affiliation{Department of Physics, The Hong Kong University of Science and Technology, Clear Water Bay, Kowloon, Hong Kong SAR}
\affiliation{These authors contributed equally}
\author{Jiangchang Zheng}
\affiliation{Department of Physics, The Hong Kong University of Science and Technology, Clear Water Bay, Kowloon, Hong Kong SAR}
\affiliation{These authors contributed equally}
\author{Ruopeng Yu}
\affiliation{Department of Physics, The Hong Kong University of Science and Technology, Clear Water Bay, Kowloon, Hong Kong SAR}
\affiliation{These authors contributed equally}
\author{Soumya Sankar}
\affiliation{Department of Physics, The Hong Kong University of Science and Technology, Clear Water Bay, Kowloon, Hong Kong SAR}
\author{Hoi Chun Po}
\affiliation{Department of Physics, The Hong Kong University of Science and Technology, Clear Water Bay, Kowloon, Hong Kong SAR}
\author{Kam Tuen Law}
\affiliation{Department of Physics, The Hong Kong University of Science and Technology, Clear Water Bay, Kowloon, Hong Kong SAR}
\author{Berthold J\"{a}ck}
\email[]{bjaeck@ust.hk}
\affiliation{Department of Physics, The Hong Kong University of Science and Technology, Clear Water Bay, Kowloon, Hong Kong SAR}

\date{\today}

\begin{abstract}
Destructive interference between electron wavefunctions on the two-dimensional (2D) kagome lattice induces an electronic flat band, which could host a variety of interesting many-body quantum states. Key to realize these proposals is to demonstrate the real space localization of kagome flat band electrons. In particular, the extent to which the often more complex lattice structure and orbital composition of realistic materials counteract the localizing effect of destructive interference, described by the 2D kagome lattice model, is hitherto unknown. We used scanning tunneling microscopy (STM) to visualize the non-trivial Wannier states of a kagome flat band at the surface of CoSn, a kagome metal. We find that the local density of states associated with the flat bands of CoSn is localized at the center of the kagome lattice, consistent with theoretical expectations for their corresponding Wannier states. Our results show that these states exhibit an extremely small localization length of two to three angstroms concomitant with a strongly renormalized quasiparticle velocity $v\approx1\times10^4\,$m/s, which is comparable to that of moiré superlattices. Hence, interaction effects in the flat bands of CoSn could be much more significant than previously thought. Our findings provide fundamental insight into the electronic properties of kagome metals and are a key step for future research on emergent many-body states in transition metal based kagome materials.

\end{abstract}

\maketitle

\begin{figure*}[htbp]
\includegraphics[width=17cm]{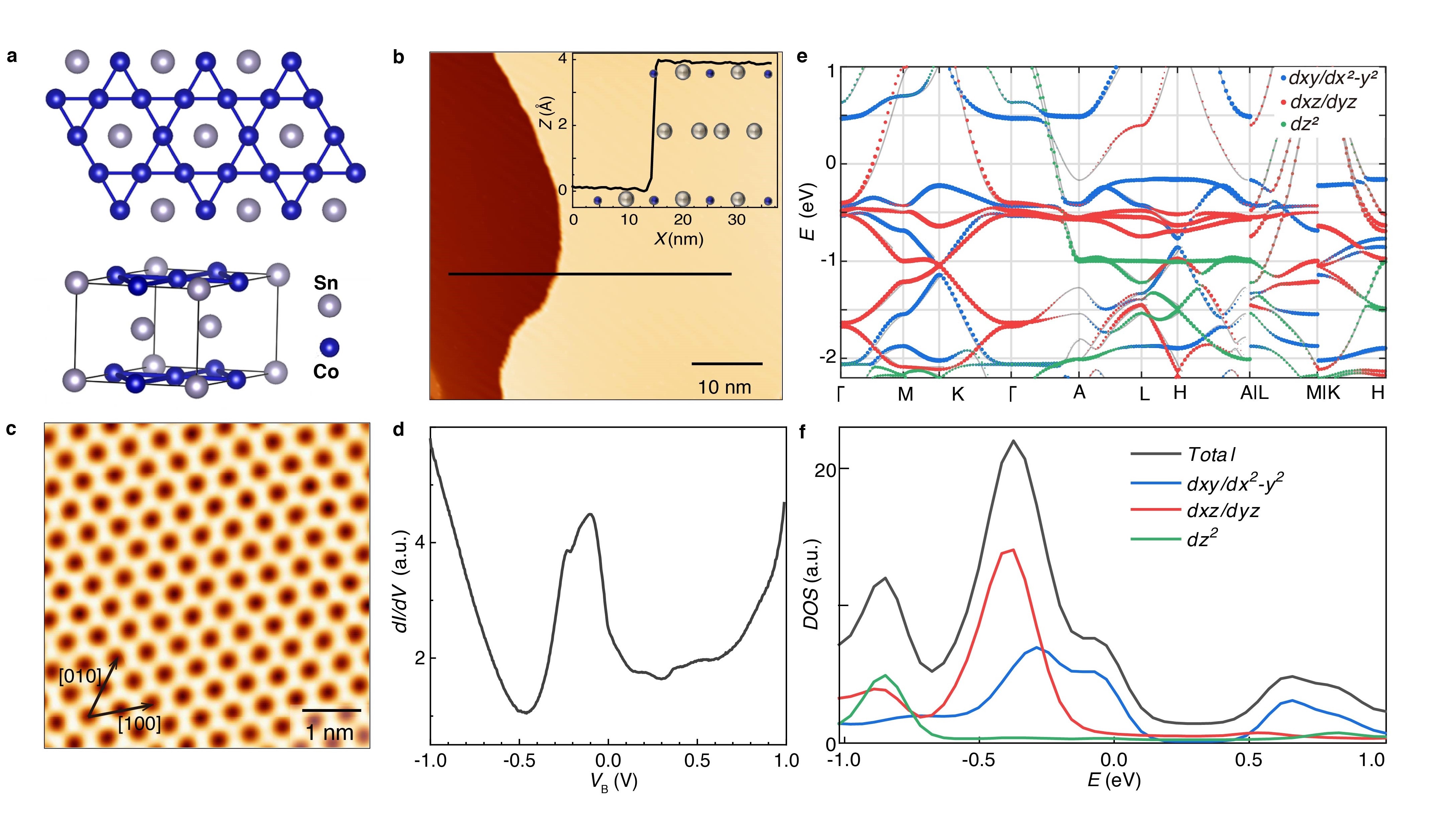}
\centering
\caption{{\bf Topographic and electronic properties of CoSn thin films.} {\bf a}, the lattice structure of CoSn. Upper panel: Lattice structure of Co$_3$Sn kagome layer within the crystallographic {\em a-b}-plane (Silver sphere: Sn; blue sphere: Co). Lower panel: Yhree-dimensional lattice structure of CoSn shows the alternating stacking of Co$_3$Sn and stanene Sn$_2$ layers along the crystallographic {\em c}-axis. {\bf b}, scanning tunneling microscopy (STM) topography conducted on the surface of the as-grown CoSn thin films ($I$= 50 pA, $V_{\rm B}$= -1.5\,V). The inset displays the topographic height recorded along the black line in the topographic image. {\bf c}, High-resolution STM topography of the Co$_3$Sn layer ($I=3\,$nA, $V_{\rm B}=0.3\,$V). {\bf d},  $dI/dV$ spectrum recorded on top of the Co$_{3}$Sn layer ($I=3\,$nA, $V_{\rm modulation}$ ($V_{\rm m}$)=10\,mV). {\bf e}, Spectral function obtained from density functional theory calculations along the crystallographic high-symmetry directions (see Methods). The orbital origin of the individual bands is encoded in the colors (see legend). {\bf f}, The calculated total and orbital-resolved density of states (DOS) of the Co $d$-orbital derived bands shown in panel e.}
\label{fig:01}
\end{figure*}

\section{I. INTRODUCTION}
Electrons occupying flat bands in momentum space are expected to be localized in real space. Hence, materials with electronic flat bands present an attractive venue to explore many-body quantum states that can arise from Coulomb interactions. To date, much interest has been directed to the study of many-body quantum states appearing in moiré flat bands~\cite{cao2018unconventional, cao2018correlated, serlin2020intrinsic, tang2020simulation, regan2020mott, xie2021fractional}. The two-dimensional kagome lattice model, realized through a set of corner sharing triangles (see Fig.~\ref{fig:01}a), offers an alternative way to realize a topological flat band~\cite{sun2011nearly, guo2009topological}. Here, destructive interference between electron wavefunctions is predicted to result in electronic states with non-trivial real space Wannier functions~\cite{kang2020topological} which are localized at atomic length scales. The experimental demonstration of these characteristics will be key toward the realization of various interesting quantum phases, such as zero-field fractional quantum Hall states~\cite{tang2011high, neupert2011fractional}, unconventional superconductivity~\cite{cao2018unconventional}, and Hubbard-Mott transitions~\cite{tang2020simulation, regan2020mott}. Moreover, a variety of kagome materials that exhibit topological electronic states~\cite{ye2018massive}, superconductivity~\cite{ortiz2020cs}, and chiral charge density waves~\cite{arachchige2022charge} have been discovered recently. Experimental insights into the electronic properties of kagome metals will generally promote a better understanding of this rich phenomenology.

The kagome metals of the FeSn family have attracted significant interest, because the judicious choice of constituent elements and stoichiometry permits control of the electronic, magnetic, and topological properties~\cite{ye2018massive, kang2020dirac, kang2020topological, liu2020orbital, xing2020localized, liu2018giant, sankar2023observation}. In this regard, CoSn was identified as a particularly promising candidate~\cite{sales2022flat} in which an advantageous layer stacking sequence combined with the absence of long-range magnetic order results in a set of weakly dispersing bands near Fermi energy ($E_{\rm F}$)~\cite{kang2020topological, liu2020orbital, meier2020flat}. It is important to appreciate that realistic materials deviate from the 2D kagome lattice model; the electronic coupling in the out-of-plane direction, a more complex lattice structure, and spin-orbit coupling could weaken the electron localization. The inherent interest in flat band systems is in the potential amplification of interaction effects between localized charge carriers. The effective interaction strength is strongly affected by the degree of localization of the relevant orbitals, and how it suffers from these deteriorating effects is not known so far. 

Here, we perform the first real-space characterization of electrons occupying a kagome flat band using STM. Such measurements are particularly suited to study the energy-resolved local electronic density of states (LDOS) at atomic resolution and previously helped characterize the flat bands of moiré superlattices~\cite{li2010observation, wong2015local, li2021imaging}. In this work, we use spectroscopic imaging with the STM on the surface of CoSn thin films to visualize the real space localization of kagome flat band electrons. Going beyond a bandwidth characterization in momentum space~\cite{kang2020topological, liu2020orbital}, the high spatial resolution of STM enables us to distinguish the non-trivial Wannier states of the kagome flat band from trivially localized states. Surprisingly, they appear more localized than what earlier theories predict~\cite{kang2020topological} and exhibit a strongly renormalized quasiparticle velocity. This suggests the enhancement of interaction effects in these flat bands could be more significant than previously thought.

\section{II. RESULTS}
CoSn ($P6/mmm$ space group, $a=5.25\,$Å, $c=4.19\,$Å) crystallizes in a hexagonal lattice structure that consists of alternating Sn$_2$ honeycomb (stanene) and Co$_3$Sn kagome layers. We prepared CoSn thin films (nominal thickness $d=\,$50 nm) on the surface of niobium (Nb) doped SrTiO$_3$ (111) by using molecular beam epitaxy (MBE) (see Methods, as well as Ref.~\cite{SI} for materials characterization). The Vollmer-Weber growth mode results in flat top islands with typical diameters of few hundreds of nanometers~\cite{SI}. STM measurements conducted on the surface of these islands show large, atomically flat, and defect-free terraces (Fig.~\ref{fig:01}b), which makes them suitable for spectroscopic imaging measurements with the STM to visualize the kagome flat bands. High-resolution STM topographies recorded on various islands show that their surface is terminated by a Co$_3$Sn layer, whose apparent height is dominated by the $d$-orbitals of the Co atoms occupying the kagome lattice sites (Fig.~\ref{fig:01}c)~\cite{liu2020orbital}. This observation is consistent with the high substrate temperature $T>800\,$°C during film deposition, which disfavors a stanene termination~\cite{zhu2015epitaxial}. A detailed analysis of the Co$_3$Sn lattice structure (Fig.~\ref{fig:01}a) further reveals the presence of approximately $2\,\%$ tensile strain along the [010]-direction (crystallographic {\em b}-axis), which presumably results from the lattice mismatch ($\approx5\,\%$) with the substrate.

\begin{figure*}%[htbp]
\includegraphics[width=17cm]{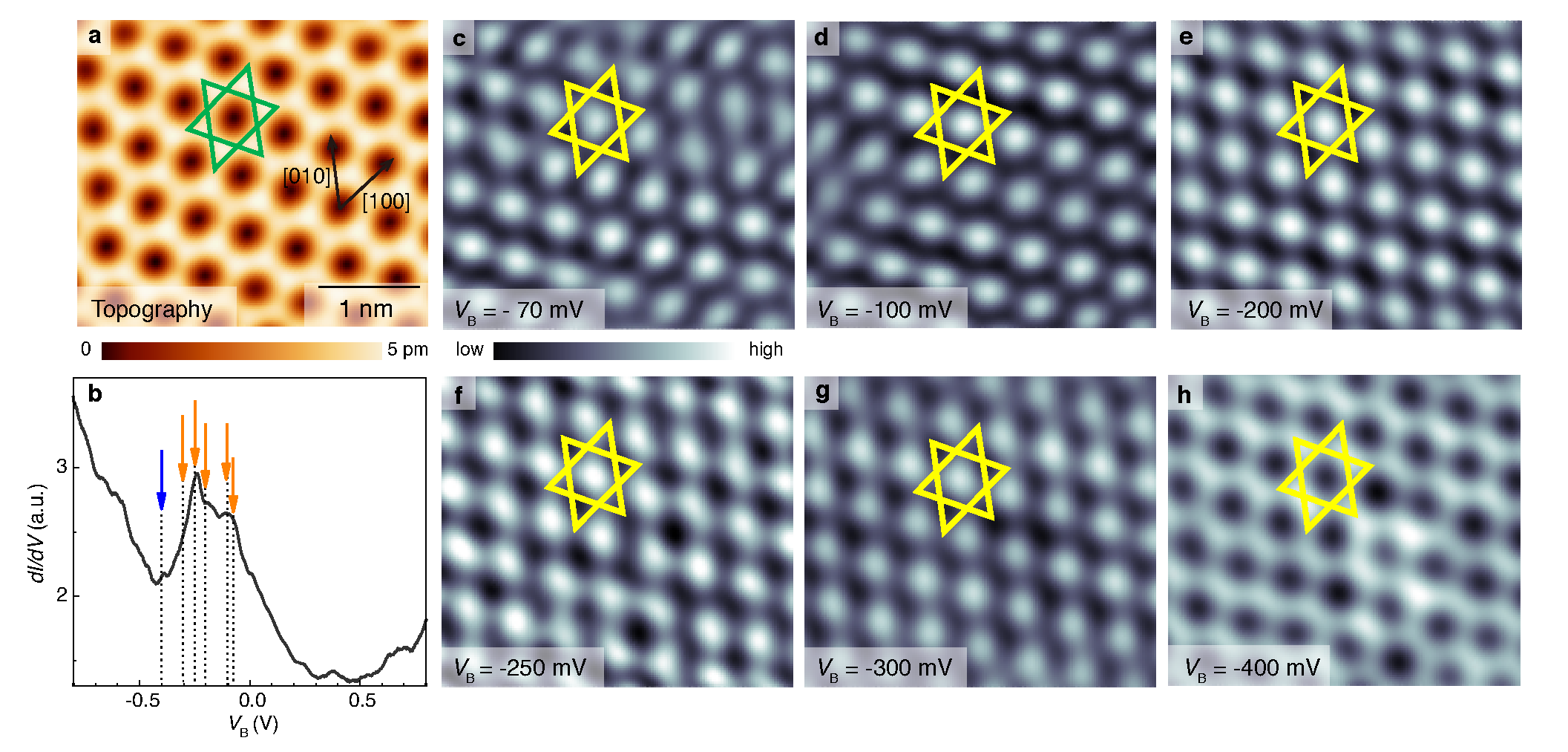}
\centering
\caption{\textbf{Visualizing the kagome flat band of CoSn}. {\bf a}, high-resolution topography recorded on the thin film surface terminated by a Co$_3$Sn layer ($I$=3.5\,nA, $V_{\rm B}$= 300 mV). {\bf b}, $dI/dV$ spectrum recorded within the field of view of panel a. {\bf c}-{\bf h}, a series of $dI/dV$ maps recorded across a range of $V_{\rm B}$ set points from -70 meV to -400 meV (see Methods) in the same field of view as panel a. The green and yellow lines indicate the position of the kagome lattice in each image ($I$= 3 nA, $V_{\rm m}$= 10\,mV). }
\label{fig:02}
\end{figure*}

\begin{figure*}[htbp]
\includegraphics[width=12.75cm]{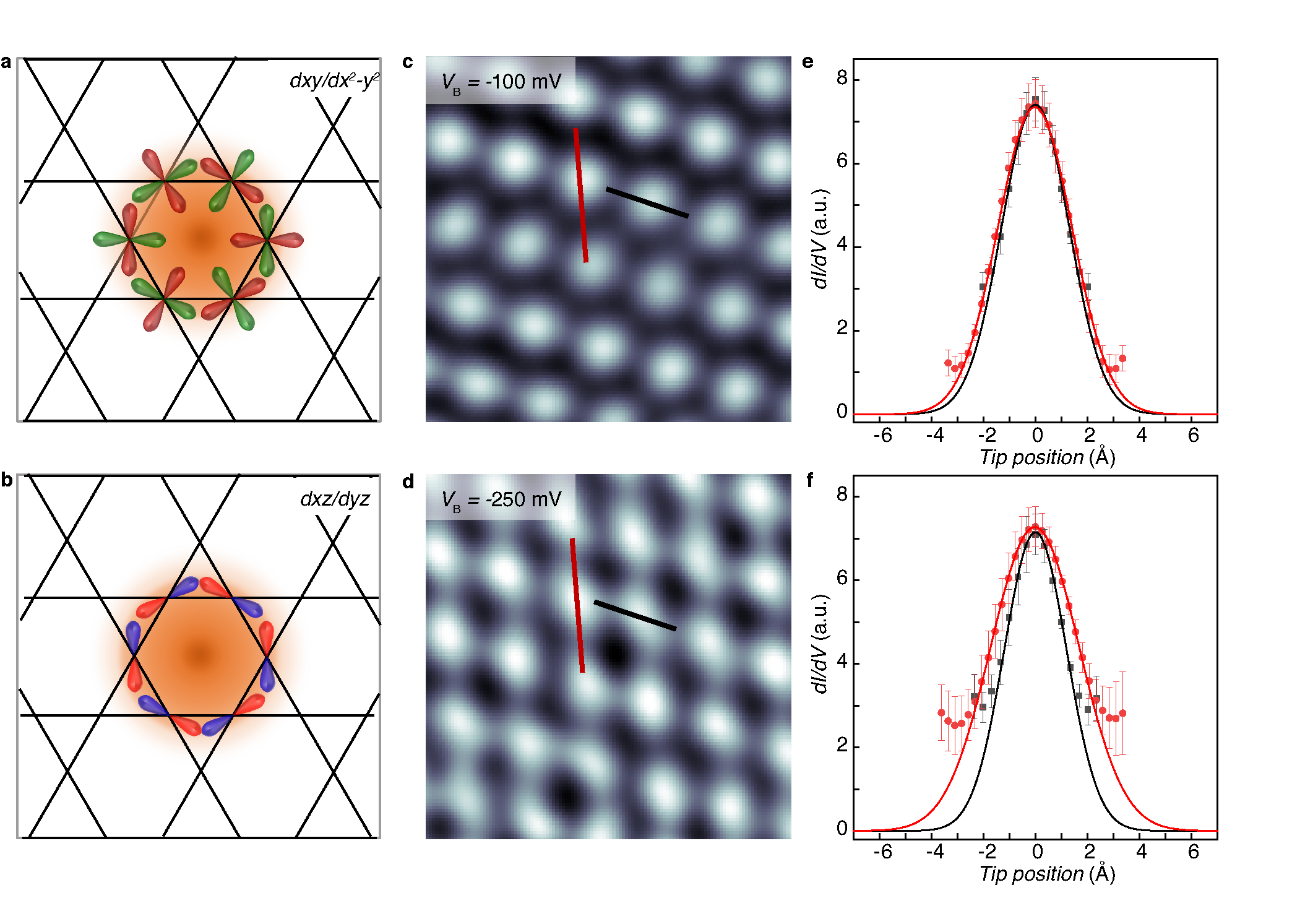}
\centering
\caption{ {\bf Localization length of flat band electrons in CoSn.} {\bf a} and {\bf b}, Spectral weight of the Wannier states (orange color) of FB1 and FB2 constructed from the $d_{\rm xy}/d_{\rm x^2-y^2}$ (red/green lobes) and  $d_{\rm xz}/d_{\rm yz}$ (red/blue lobes) orbitals, respectively. The These panels were adopted from~\cite{kang2020topological}. {\bf c} and {\bf d}, $dI/dV$ maps recorded at $V_{\rm B}=-100\,$mV and $V_{\rm B}=-250\,$mV ($I$= 3 nA, $V_{\rm m}$= 10\,mV). {\bf e} and {\bf f}, $dI/dV$ amplitudes (solid symbols) extracted from the spectroscopic maps along line profiles shown in panels {\bf c} and {\bf d}, respectively. Gaussian fits to the data are shown as solid lines in the respective colors.}
\label{fig:03}
\end{figure*}

Electronic flat bands manifest as a sharp peak in the differential conductance ($dI/dV$) spectrum~\cite{li2010observation, wong2015local, xie2019spectroscopic} of scanning tunneling spectroscopy (STS) measurements, owing to the flat band's large LDOS ($\propto dI/dV$) within a small energy window. We have performed STS measurements to detect the presence of such a spectral feature at the surface of CoSn. The $dI/dV$ spectrum (Fig.~\ref{fig:01}d) is dominated by a prominent double peak structure, which is located just below zero applied bias voltage $V_{\rm B}$. Comparing these characteristics with the electronic structure and LDOS obtained from {\em ab-initio} calculations (Fig.~\ref{fig:01}, d and f, and Methods) and previous results from photo-electron spectroscopy measurements~\cite{liu2020orbital, kang2020topological}, these peaks can be associated with the LDOS of the kagome flat bands. These calculations reveal that two flat bands are situated near Fermi energy $E_{\rm F}$ and derive from the Co $3d$-orbitals (Fig.~\ref{fig:01}, e and f); The first flat band (FB1), which is located just below $E_{\rm F}$, is formed by $d_{\rm xy}$ and $d_{\rm x^2-y^2}$ orbitals while a second flat band (FB2), appearing at lower energies, originates from the $d_{\rm xz}$ and $d_{\rm yz}$ orbitals. This observation of a pronounced $dI/dV$ peak associated with the CoSn flat bands contrasts with our results from STM studies on thin films of the iso-structural antiferromagnet FeSn~\cite{SI}. Here, $dI/dV$ spectra recorded on the Fe$_3$Sn surface lack this distinct feature~\cite{lee2022spin, li2022spin, multer2023imaging}, because the electronic structure of FeSn does not have a kagome flat band, owing to the deteriorating influence of strong magnetic exchange terms (see Ref.~\cite{SI}). We further note that the tensile strain along the [010]-direction slightly modifies the spectral appearance of the $dI/dV$ peaks associated with the flat band LDOS but it does not significantly affect neither the height nor the width of this spectral feature (see Fig.~\ref{fig:02}, Fig.~\ref{fig:03} and Ref.~\cite{SI} for measurements on other CoSn islands).

\begin{figure*}[htbp]
\includegraphics[width=16.75cm]{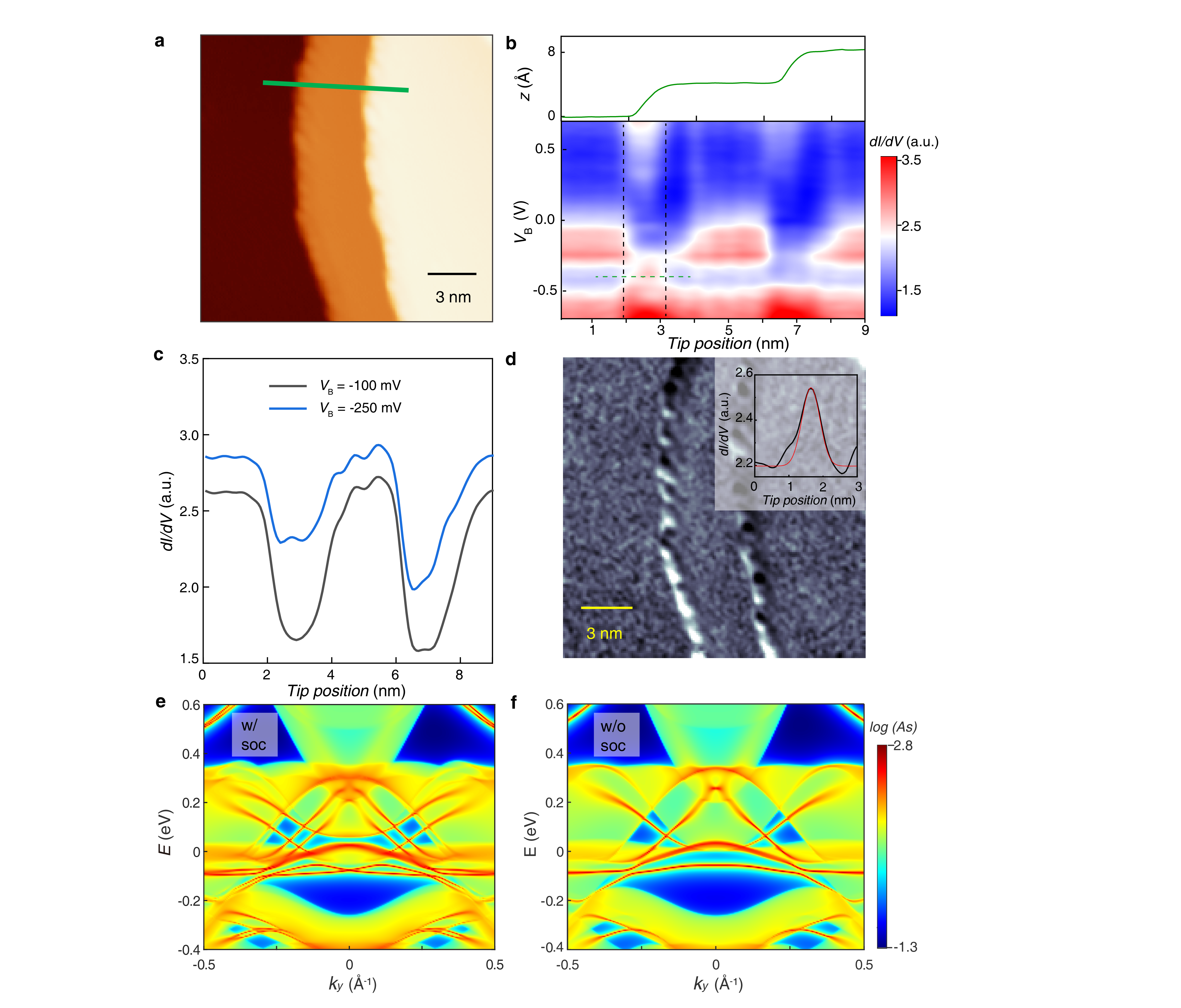}
\centering
\caption{\textbf{Characterizing the flat band properties at atomic step edges.} {\bf a}, STM topography of the CoSn surface, which exhibits two one unit cell step edges ($I=2.5\,$nA, $V_{\rm B}=400\,$mV). {\bf b}, Upper panel: The topographic height $z$ profile extracted along the green line in panel~a. Bottom panel: $dI/dV$ spectra recorded along along the green line in panel~a plotted as a function of tip position and applied bias voltage $V_{\rm B}$ ($I=2\,$nA, $V_{\rm m}$= 10\,mV, stabilization voltage $V_{\rm B}=-900\,$mV). The edge region is indicated by the vertical pair of dahsed lines. We note that the spatial modulation of the flat band $dI/dV$ is only faintly visible, because it is obscured by the set-point effect present in these spectroscopic line-cut measurements. \textbf{c}, $dI/dV$ amplitude recorded at $V_{\rm B}=-100\,$mV and $V_{\rm B}=-250\,$mV plotted as a function of the tip position along the green line in panel \textbf{a} ($I$= 2 nA, $V_{\rm m}$= 10\,mV). {\bf d}, Spatially mapped $dI/dV$ amplitude recorded in the field of view of panel~a at $V_{\rm B}=-350\,$mV. The inset displays the position-dependent $dI/dV$ amplitude at $V_{\rm B}=-350\,$mV recorded along the horizontal green dashed line in panel {\bf b}. {\bf e} and {\bf f}, the spectral functions ($As$), which were calculated from a Wannier tight binding model, display the energy ($E$) and momentum ($k_{\rm y}$) dependent spectral weight at the edge of a CoSn monolayer (one unit cell thickness, 40 unit cells width) with and without spin-orbit coupling (SOC), respectively. See Methods for model details.}
\label{fig:04}
\end{figure*}

The Wannier functions of $d$-orbital-derived kagome flat bands are expected to be localized at the kagome lattice center~\cite{kang2020topological} ({\em cf.}~Fig.\,~\ref{fig:01}a). Our experimental results from spectroscopic imaging measurements are consistent with this expectation (see Ref.~\cite{SI} for results on another CoSn island). The spectroscopic map recorded at a bias voltage $V_{\rm B}$= -400 mV (Fig.~\ref{fig:02}h), which corresponds to an energy outside the flat band $dI/dV$ peak mirrors the corresponding STM topography (Fig.~\ref{fig:02}a). It reflects the Co atom positions in the kagome lattice of the Co$_3$Sn layer, whose $d$-orbitals contribute the spectral weight of the dispersive electron bands at this energy $E=eV_{\rm B}$ ({\em cf.}~Fig.~\ref{fig:01},a and e; $e$ denotes the electron charge). Interestingly, spectroscopic maps recorded within the flat band $dI/dV$ peak at -300\,mV$\leqslant V_{\rm B}\leqslant-70$ mV exhibit a contrast reversal; an enhanced $dI/dV$ amplitude is observed at the kagome lattice center, which is not occupied by a Co atom (Fig.~\ref{fig:02}, c to g). We note the Sn atom at the kagome lattice center of the Co$_3$Sn layer ({\em cf.}~Fig.~\ref{fig:01}a) contributes negligible spectral weight at these energies and cannot account for the observed $dI/dV$ characteristics~\cite{SI}. Hence, this observation serves as direct evidence for the non-trivial character of the flat band Wannier states, and it allows us to distinguish them from trivially localized states, which would reside at the kagome lattice sites of the Co atoms. By contrast, the real-space identification of the non-trivial Wannier state of a kagome flat band would not be easily feasible in case of {\em s}- and {\em p}-orbital derived flat bands, whose spectral weight is expected to reside on the kagome lattice sites~\cite{li2018realization}. Interestingly, the localized LDOS in spectroscopic maps recorded at $V_{\rm B} \geqslant$-100\,mV (Fig.~\ref{fig:02},c and d) exhibits an circular shaped real space pattern, whereas it appears rather elongated along the [010]-direction in maps recorded at $V_{\rm B}= -250\,$mV and $V_{\rm B}=-300\,$mV (Fig.~\ref{fig:02},f and g). This different strain response indicates the different orbital origin of FB1 and FB2, consistent with theoretical expectations ({\em cf.} Fig.~\ref{fig:01}e). 

Previous theoretical analyses of the Wannier states of the CoSn flat bands predict a  localization length $l\approx7\,$Å~\cite{kang2020topological}. Interestingly, our quantitative analysis of the spatial $dI/dV$ distributions reveals much shorter values. We analyze $l$ as the spatial decay of the $dI/dV$ amplitude and apply Gaussian fits to representative $dI/dV$ line cuts (see Fig.~\ref{fig:03}, c to f). The $d_{\rm xy}/d_{\rm x^2-y^2}$-derived FB1 exhibits localization lengths $l_{\rm FB1}=2.6\pm 0.1\,$Å and $l_{\rm FB1'}=2.9\pm 0.1\,$Å along the [100]- and [010]-directions, respectively. Similarly, the $d_{\rm xz}/d_{\rm yz}$-derived FB2 exhibits localization lengths $l_{\rm FB2}=2.4\pm 0.2\,$Å and $l_{\rm FB2'}=3.4\pm 0.5\,$Å along the [100]- and [010]-directions, respectively. While tensile strain along the [010]-direction increases the localization length along this direction, the spectral weight remains predominantly localized at the kagome lattice center (nominal diameter $\approx7.4\,$Å).

This observed real space localization of the flat band electrons suggests a moderate out-of-plane (crystallographic {\em c}-axis) coupling between adjacent Co$_3$Sn layers, that is, the kagome flat band of CoSn should be quasi-two-dimensional. Unlike three-dimensional electronic states which extend into the bulk, 2D electronic states are strongly perturbed by surface defects, such as atoms~\cite{heller1994scattering} and step edges~\cite{drozdov2014one}. We test the dimensionality of the CoSn flat bands by recording $dI/dV$ spectra along a line, which crosses two one unit cell high step edges (Fig.~\ref{fig:04}a). The $dI/dV$ amplitude associated with the dispersive Co bands at $V_{\rm B}\leqslant-400\,$mV weakly responds to the presence of the step edges (Fig.~\ref{fig:04}b). This is consistent with the observed out-of-plane dispersion of these electronic states in from a previous study~\cite{kang2020topological} and indicates their three-dimensional character. By contrast, the flat band $dI/dV$ amplitude at $-350\,\rm{mV}\leqslant V_{\rm B}\leqslant-70\,\rm{mV}$ exhibits a strong response to the presence of step edges and significantly decays over a distance of less than one nanometer (Fig.~\ref{fig:04}, b and c). This observation indicates the quasi-2D character of the kagome flat bands of CoSn, consistent with a vanishing out-of-plane dispersion of these bands in momentum space~\cite{kang2020topological, liu2020orbital}.

The suppression of the flat band LDOS near atomic step edges permits the presence of a localized edge state in the same energy range (Fig.~\ref{fig:04}b). Spatial mapping of the $dI/dV$ amplitude at $V_{\rm B}\approx350\,$mV confirms the one-dimensional edge state character of this spectral feature (Fig.~\ref{fig:04}d) the non-uniform distribution of which suggests that its presence depends on details of the edge termination. Analyzing the characteristic decay length $\zeta$ of this state into the quasi-2D bulk, we can estimate the quasiparticle velocity $v$ of the kagome flat bands. We extract $\zeta\approx5\,$Å by fitting the spatial decay of the edge state by using a Gaussian function (Fig.~\ref{fig:04}d inset) and estimate $v$ of the kagome flat bands via $v=(\zeta\Delta_{\rm SOC})/h$ ($h$ denotes Planck's constant). Previous measurements of the electronic structure of CoSn detected a spin-orbit coupling (SOC) induced spectral gap $\Delta_{\rm SOC}\approx80\,$meV at the $\Gamma$-point between -300\,meV and -400\,meV~\cite{kang2020topological}. Using these parameters, we obtain $v\approx1\times10^{4}\,$m/s, an extremely small value which is comparable to the quasiparticle velocity found in moiré superlattices (e.g., $v\approx4\times10^4\,$m/s for magic angle twisted bilayer graphene~\cite{cao2018correlated}. This value is also quantitatively comparable to the theoretical estimate $v^{*}\approx5\times10^{3}\,$m/s in a simple tight-binding approximation $E\approx t\cos(ka)$ ($k$ - crystal momentum) for a calculated hopping parameter $t\approx15\,$meV between the flat band Wannier states~\cite{kang2020topological}. 

Finally, we comment on the discussed topological character of the CoSn flat band~\cite{kang2020topological}. Interestingly, a SOC-induced spectral gap at the $\Gamma$-point imbues the kagome flat band in the 2D limit with a non-trivial $\mathbbm{Z}_2=1$ index. 
Our tight-binding model calculations (see Methods and Ref.~\cite{SI}) of one CoSn layer (1 unit cell thickness, 40-unit cells width) indeed show the presence of 1D edge states, which appear at the flat band energy. They connect the flat bands with the dispersive conduction and valence bands of the 2D bulk over a large energy range (Fig.~\ref{fig:04}, e and f). However, these states appear both in case SOC is included and excluded and SOC merely induces a spectral splitting. The opening of a non-trivial $\Delta_{\rm SOC, DFT}\approx50\,$meV due to SOC~\cite{kang2020topological} (see ref.~\cite{SI} for the bulk spectral functions) could indeed reconnect these edge states with the bulk bands in a non-trivial fashion to accommodate the $\mathbbm{Z}_2=1$ index. While we cannot resolve these details in our calculations, owing to the coupling of the edge states to the metallic bulk states, it will be interesting to conduct further experiments with magnetic scattering centers~\cite{jack2020observation} to clarify the topological origin of the observed edge states.

\section{III. DISCUSSION AND CONCLUSION}
Our experimental results demonstrate the real space localization of electrons occupying a kagome flat band. Spectroscopic mapping with the STM allowed us to directly visualize the corresponding non-trivial Wannier states, which results from destructive electron interference. While interlayer and spin-orbit coupling, and a more complex lattice structure appearing under realistic conditions are generally believed to counteract the localization mechanism of destructive interference described in the 2D kagome lattice model, our observations demonstrate that kagome flat band electrons can retain their localized nature despite these effects. This key result also emphasizes the importance of a suitable layer stacking sequence, such as found in CoSn, and may inform future material design efforts to identify other promising kagome flat band materials~\cite{meier2020flat}. 

Our quantitative analyses further show that the flat band Wannier states of CoSn are localized with $l\approx2-3\,$Å, a value much smaller than theoretically expected~\cite{kang2020topological}. This strong localization is concomitant with a significantly renormalized $v\approx1\times10^4\,$m/s. Our findings contrast with moiré superlattice materials where the larger spatial extent (e.g., $\approx 15\,$nm for magic angle twisted bilayer graphene) of the flat band Wannier functions~\cite{po2018origin} affords a comparably small Coulomb on-site repulsion $U\approx20\,$meV~\cite{cao2018correlated,xie2019spectroscopic}. Hence, the enhancement of interaction effects in the flat bands of CoSn due to a large $U\approx5-6\,$eV~\cite{kang2020topological} could be even more pronounced than previously thought~\cite{huang2022flat}. 

Together, our observations establish transition-metal based kagome metals as a promising venue to explore the emergence of interacting many-body quantum states in a topological flat band at potentially elevated temperatures~\cite{sun2011nearly,tang2011high,guo2009topological}. 
We anticipate that chemical~\cite{sales2021tuning} and modulation doping~\cite{cheng2022atomic}, as well as lattice strain~\cite{kang2020topological} (see Ref.~\cite{SI} for a discussion of tensile strain observed in our study) offer rich opportunities to realize partial flat band occupations at which these effects are expected to appear.

\section{ACKNOWLEDGEMENTS}
This work has been primarily supported by an early career grant of the Hong Kong RGC through grant No. 26304221 and the Croucher foundation through grant No. CIA22SC02 awarded to B.J. K.T.L. acknowledges the support of the Ministry of Science and Technology, China, and Hong Kong Research Grant Council through grants Nos. 2020YFA0309600, RFS2021-6S03, C6025-19G, AoE/P-701/20, 16310520, 16310219, and 16307622. H.C.P. acknowledges support from the Hong Kong RGC through grant No. 26308021 and the Croucher Foundation through grant No. CF21SC01. C.C. acknowledges support through a postdoctoral fellowship from the Tin Ka Ping foundation.

C.C., J.Z., and S.S. prepared and characterized the samples. C.C. and J.Z. performed the STM measurements and analyzed the data. R.Y. performed the model calculations. All authors discussed the results and contributed to the manuscript, which was written by C.C. and B.J.

\section{APPENDIX: Methods}
\subsection{1. Sample preparation}
The experiments were performed in a home-built ultra-high vacuum (UHV) MBE-STM system. CoSn thin films were synthesized on single crystal Nb-doped SrTiO$_3$(111) substrates (CrysTec, $5\times5\times0.5$ mm, 0.05 wt\% Nb concentration). Prior to the film synthesis, the substrates were prepared by hydrofluoric acid etching and annealed at 1050\,°C in an oxygen atmosphere for one hour to prepare atomically flat and nominally oxygen-vacancy-free surfaces. The prepared substrates were loaded into the MBE chamber and annealed in UHV at 600\,°C and 850\,°C for 60\,min and 5\,min, respectively to remove residual moisture and surface adsorbates. CoSn films were deposited by co-evaporating Co and Sn from solid source effusion cells at a background pressure $p<1\times10^{-9}\,$mbar. During the growth, the substrate was kept at a constant temperature ($\sim$ 830\,°C) and enclosed by a cryoshroud nominally held at $T=77\,$K. The ratio of beam-equivalent pressures (BEPs) was $P_{\rm Co}:P_{\rm Sn}$=1:4.3, where $P_{\rm Co}$ and $P_{\rm Sn}$ denote the BEPs of Co and Sn, respectively. The nominal thicknesses of the grown films are $\approx40-50\,$nm.
\subsection{2. STM measurements}
The STM and STS measurements were performed at liquid helium temperature and UHV conditions ($p\leq1.4\times10^{-10}\,$mbar) using a chemically etched tungsten STM tip. The tip was prepared on a Cu(111) surface by field emission and controlled indention and calibrated against the Cu(111) Shockley surface state before each set of measurements. $dI/dV$ spectra were recorded using standard lock-in techniques with a small bias modulation $V_{\rm m}$ chosen between $V_{\rm m}=1\,$mV to $V_{\rm m}=10\,$mV at a frequency $f=971\,$Hz. $dI/dV$ maps were recorded using the multi-pass mode. Here, the topographic height profile is first recorded in constant current mode at a suitable bias voltage. In the next step, the recorded topographic profile is replayed under open feedback conditions and the position-dependent $dI/dV$ amplitude is recorded at a chosen $V_{\rm B}$ in a set of multiple passes along the same topographic line. This allows for the acquisition of high-quality $dI/dV$ maps while avoiding set-points effects.
\subsection{3. Model calculations}
Density functional theory (DFT) calculations were performed using the Vienna Ab-Initio Simulation Package (VASP)~\cite{kresse1996efficient, kresse1996efficiency}. The pseudo-potential is from the projector-augmented wave (PAW) method~\cite{blochl1994projector}, with the exchange-correlation functional given by the generalized-gradient approximation (GGA) parameterized by Perdew-Berke-Ernzerhof (PBE)~\cite{perdew1996generalized}.The plane-wave cut-off energy is set to 300\,eV, and a $k$-point mesh of $9\times9\times9$ is adopted for the first Brillouin zone sampling. The Wannier tight-binding Hamiltonian of bulk is constructed using the Wannier90 package~\cite{mostofi2014updated}; the  $d$-orbitals of the Co atoms and the $p$-orbitals of the Sn atoms are selected as the initial projection basis. Based on this tight-binding model of bulk CoSn, the effective Hamiltonian of the geometry with certain real space truncation can be constructed.

\bibliography{bibliography}
\end{document}